# Reflectionless Sharp Bends and Corners in Waveguides Using Epsilon-Near-Zero Effects

Brian Edwards[1], Andrea Alù[1], Mário G. Silveirinha[1,2], Nader Engheta[1*]

*(1) Department of Electrical and Systems Engineering, University of Pennsylvania, Philadelphia, PA 19104 USA*

*(2) Universidade de Coimbra, Electrical Engineering Department – Instituto de Telecomunicações, Portugal*

*Following our recent theoretical and experimental results that show how zero-permittivity metamaterials may provide anomalous tunneling and energy squeezing through ultranarrow waveguide channels, here we report an experimental investigation of the bending features relative to this counterintuitive resonant effect. We generate the required effectively-zero permittivity using a waveguide operating at the cut-off of its dominant mode, and we show how sharp and narrow bends may be inserted within the propagation channel without causing any sensible reflection or loss.*

PACS Numbers: 78.66.Sq, 42.82.Et, 52.40.Db, 52.40.Fd

---

[*] To whom correspondence should be addressed.  E-mail: engheta@ee.upenn.edu

1. Introduction

Metamaterials with near-zero permittivity (epsilon-near-zero, or ENZ) have received increased attention in the last few years for several anomalous properties that characterize their wave interaction [1]-[8]. These effects may play an important role in transparency and cloaking phenomena, pattern reshaping, antennas and nanoantennas, nanocircuits, energy squeezing and supercoupling.

Typically, metamaterials and artificial media may be well described by a "homogenized" effective permittivity and permeability following the same principles that allow any material, inherently a complicated arrangement of atoms and molecules, to be compactly described by its "bulk" properties. However, within metamaterials there is often a second level of homogenization, consisting of an artificial intermediate structure whose characteristic size is much greater than the scale of the atoms, but much less than that of the wavelength. The elements of this ordered structure are usually compact resonators that strongly interact with the electromagnetic field, providing anomalous values of effective permittivity and permeability that may not be naturally available. As an alternative realization, one may make use of the dispersive properties of electromagnetic waveguides, substituting the wave interaction of small-scale resonators with the dispersion relations produced by the waveguide walls [9]-[13]. For instance, the effective permittivity "experienced" by the dominant transverse-electric (TE$_{10}$) mode in a rectangular metallic waveguide is:

$$\varepsilon_{\text{eff}} / \varepsilon_0 = n^2 - c^2 / \left(4 f^2 w^2\right), \tag{1}$$

where $w$ is the waveguide H-plane width, $n$ is the relative refractive index of the uniform dielectric filling the waveguide, $c$ is the speed of light in vacuum, and $f$ is the operating frequency [10]. For decades, arrays of such waveguides have been employed to realize artificial materials and lenses [10] with low or negative effective permittivity, and recently the inherently negative effective permittivity provided by a single waveguide below cut-off has been exploited in a variety of applications involving metamaterials [9]-[13]. In [9], in particular, we have applied the effective zero permittivity arising at the waveguide cut-off (at frequency $f = c/2wn$) to verify experimentally the supercoupling effect and energy squeezing in ultranarrow ENZ channels, as theoretically predicted in [6]-[7].

Indeed, as was theoretically predicted in [6]-[7] and experimentally verified in [8]-[9], due to the "static-like" character of the fields in ENZ materials, it may be possible to squeeze and tunnel a significant amount of electromagnetic energy through a very narrow ENZ-filled channel. When such a channel connects two much larger waveguide sections, an anomalous resonance may enable the supercoupling phenomenon, ensuring ultralow phase variation, independent of the length of the ENZ channel and of its specific cross-sectional geometry (as long as the longitudinal cross sectional area of the transition region is kept sufficiently small). This means that the channel may make arbitrarily sharp turns and be of arbitrary length without affecting its transmission properties. For a straight channel, we have extensively proven these concepts experimentally using a U-shaped microwave waveguide, in which an ultranarrow channel operated around its cut-off frequency [9] (and thus behaving effectively as an ENZ material) was connecting two thicker waveguides operated well above their cut-off. Similar experimental proofs have been obtained in a sharp 180 [deg] microstrip bend using resonant complementary split-ring resonator particles to synthesize the ENZ metamaterial [8]. In the following, we present our recent experimental and numerical results related to the addition of sharp and arbitrary bends in the narrow waveguide channel "suprcoupling" two much larger waveguide sections. As in our previous work [9], here we employ a channel at cut-off to mimic the long wavelength characteristics of ENZ materials. We underline that the advantage of using this technique to realize the required ENZ effect, as compared with using inclusions of properly designed resonators (see e.g. [8]), does not only reside in the inherent simplicity of the design, but also in the avoidance of major causes of losses, and effects of disorder and imperfections in the experimental realization.

### 2. Experimental Setup and Realization

Our experimental setup contains three sections as shown in Fig. 1, consisting of an ultranarrow channel sandwiched between two copper coated Teflon ($\varepsilon = 2.0\varepsilon_0$) waveguides. Both of these waveguides have a height $h = 50.8\,mm$ and an H-plane width $w = 101.4\,mm$, designed so that the fundamental $TE_{10}$ mode has its cut-off at frequency $f = 1.04\,GHz$, with monomodal propagation until $2.09\,GHz$. The narrow channel has the same H-plane width, but is filled with air, ensuring a higher cut-off frequency at $f_0 = 1.48\,GHz$. Three probes have been inserted into each waveguide to extract the reflection from and the transmission through the narrow channel connecting them. Within the left waveguide, probe 1 is used to generate the $TE_{10}$ mode, while the smaller probes 2 and 3 are used to sample the local electric field. The signals collected at 2 and 3 are then used to reconstruct the impinging and reflected modes referenced to plane A. Within the second waveguide, probes 4 and 5 perform the same function, de-embedding the sampled signals into transmitted and reflected modes referenced to plane B. Alternatively and symmetrically, the wave may be excited from the other side of the setup at probe 6. These eight signals (four obtained when the excitation is at probe 1 and four when it is at probe 6) can then be used to determine the four scattering parameters of the channel. All measurements were performed using an Agilent ENA 5071C vector network analyzer. This setup and reconstruction procedure is analogous to our previous experimental setup presented in [9], where the narrow channel was straight and co-planar with the connected waveguide sections. Here, however, we concentrate on the possibility of arbitrarily bending the channel, as theoretically suggested in [7]. Moreover, we have also improved the realization of the experimental setup, eliminating some imperfections in the connection between the different parts, which caused non-negligible losses.

### 3. Results and Discussion

A total of ten different channels have been designed, constructed and tested, five examining 90° bends and five examining 180° bends. For both geometries, we have analyzed the effects of varying the length $L$ and the height $a$ of the channel. Additionally, full-wave simulations were performed to validate the experimental measurements.

The measured and simulated transmission and reflection coefficients are reported in Fig. 2 for the 180° bend and in Fig. 3 for the 90° bend. It is noticed that indeed, regardless of the geometry (90° bend or 180° bend), the length $L$, and the height $a$, there is consistently an amplitude peak in the transmission coefficient and a substantial drop in the reflection at the cut-off frequency $f_0$ of the channel, in its ENZ operation. Remarkably different from the Fabry-Perot transmission resonances seen at higher frequencies, which are predictably shifted by a change in the channel length, and possibly also by its geometry, the supercoupling resonant frequency is not sensibly shifted by such variations. Changes in the height $a$ affect only the bandwidth of the transmission peaks, with smaller heights yielding narrower peaks, due to the higher resonance Q factor. These results are consistent with the straight channel analyzed in [9], showing how the ENZ supercoupling effect and energy squeezing is not affected by bending or shape variations in the connecting channel, since it relies on a supercoupling resonance for which the effective wavelength in the channel is extremely large. Also the phase distribution confirms the ENZ response of the narrow waveguide channel at cutoff: regardless of the geometry, length $L$, or height $a$, the phase difference between the entrance and exit planes is always near zero.

Further evidence of the near zero phase variation through the ENZ region is given through the simulation results of this geometry, reported in Fig 4 for the 180[deg] bend and Fig. 5 for the 90[deg] bend. A plot of the Poynting vector is also reported, demonstrating how the energy squeezes into a region with a height $3/32^{nd}$ of the Teflon waveguide. This channel height was used for its clarity in the figure. Experiments were performed with even lower values, to yield a ratio between the heights of 1:32. Consistently with what expected theoretically [6-7], these narrower channels indeed yield even smaller phase variation.

As a final example of the robustness and flexibility of the ENZ operation, we report in Fig. 6 a numerical simulation for a more dramatic example of a bent channel, wherein the letters "ENZ" are spelled out along the ultranarrow channel. Due to its length and irregularities, there are many Fabry-Perot peaks between 1.1 GHz and 2.0 GHz.

However, the field profile shown in the figure, corresponding to the same ENZ frequency $f_0$ as in the previous examples, demonstrates unity transmission and near zero phase difference between the entrance and exit faces, despite the complex bending, channel profile and its large cross-section mismatch compared to the outer waveguide sections. The presented results support the versatility of the ENZ concepts and their possible application in complicated routing and supercoupling effects.

## 4. Conclusions

In this work we have experimentally demonstrated that a narrow ENZ channel as described in [6]-[7] can indeed provide supercoupling abilities in spite of sharp bends both at the mid-plane of the channel and at the entrance and exit faces. The ENZ nature of the channel has been realized by taking advantage of the natural dispersive properties of a rectangular waveguide. This has provided unity transmission at the cut-off frequency of the channel sharp bends and both variations in the channel length and width.

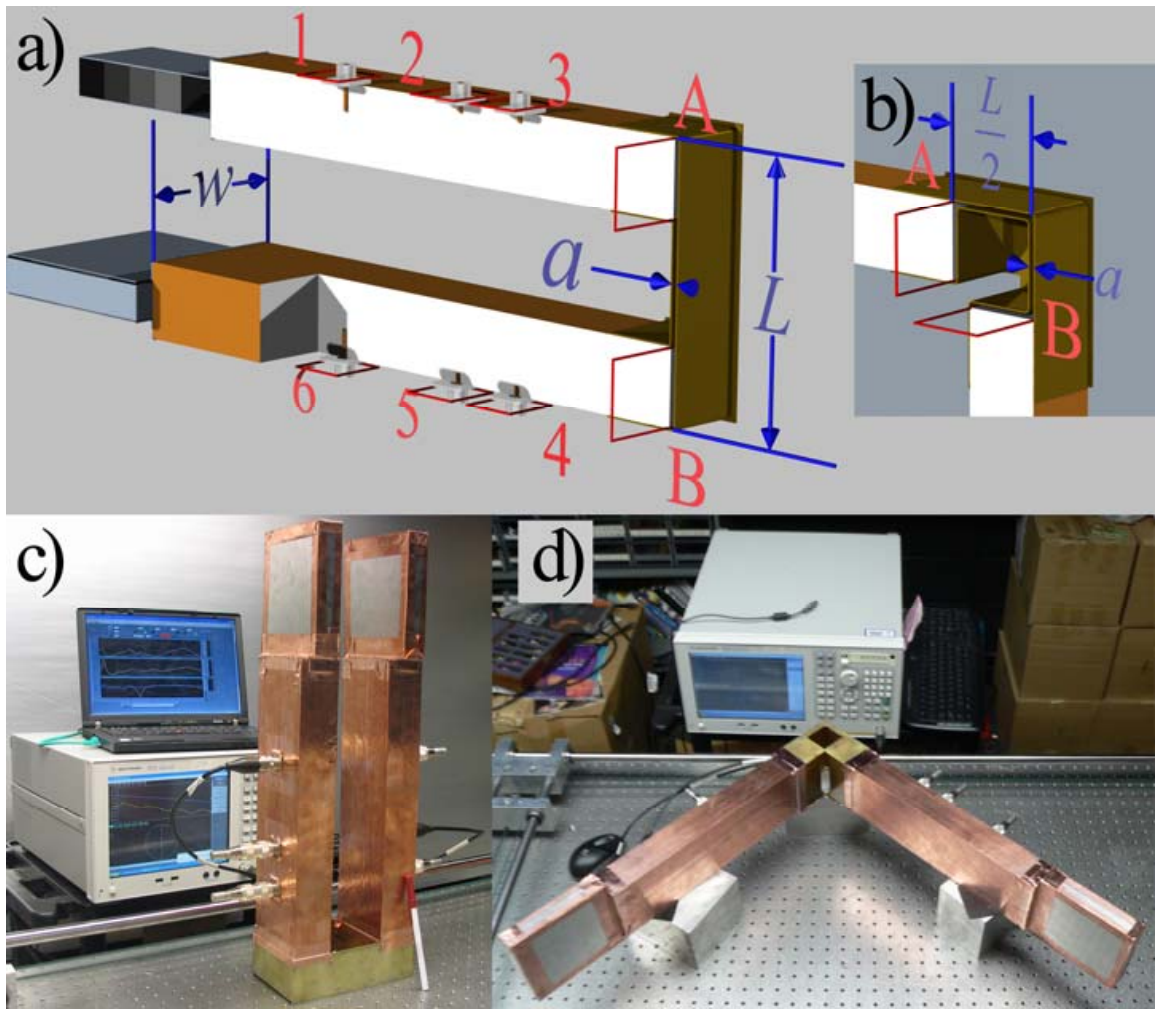

Figure 1: Cut-away schematic showing the experimental geometry for the 180º bend (panel (a)) with 90º bend in inset (panel (b)), consisting of an ultranarrow channel connecting two much thicker Teflon waveguide sections and the system of probes employed to perform the measurement. Photos of one particular geometry for the 180º bend and the 90º bend are shown in panels (c) and (d), respectively.

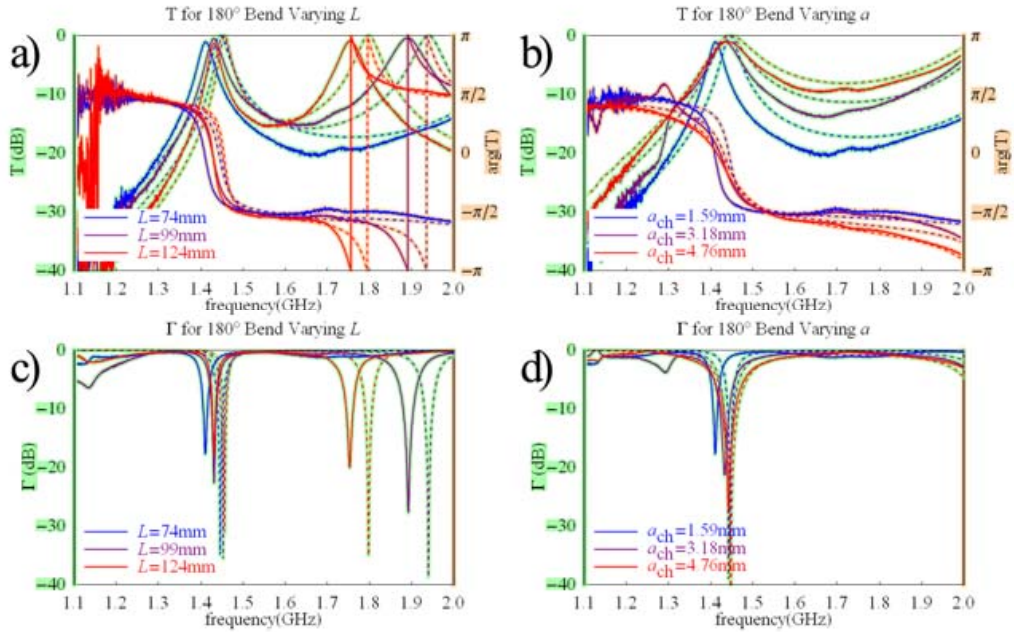

Figure 2: Experimental (solid) and simulation (dashed) results for the 180° bend. Transmission (panels (a) and (b)) and reflection (panels (c) and (d)) measurements, varying the length $L$ of the ENZ channel (panels (a) and (c)) and the height $a$ of the channel (panels (b) and (d)). Regardless of geometry, there is constantly a peak when the narrow channel is at cutoff, and behaves effectively as an ENZ material. The phase variation across the channel (panels (a) and (b)) at this frequency is always very small, confirming the ENZ operation of the channel.

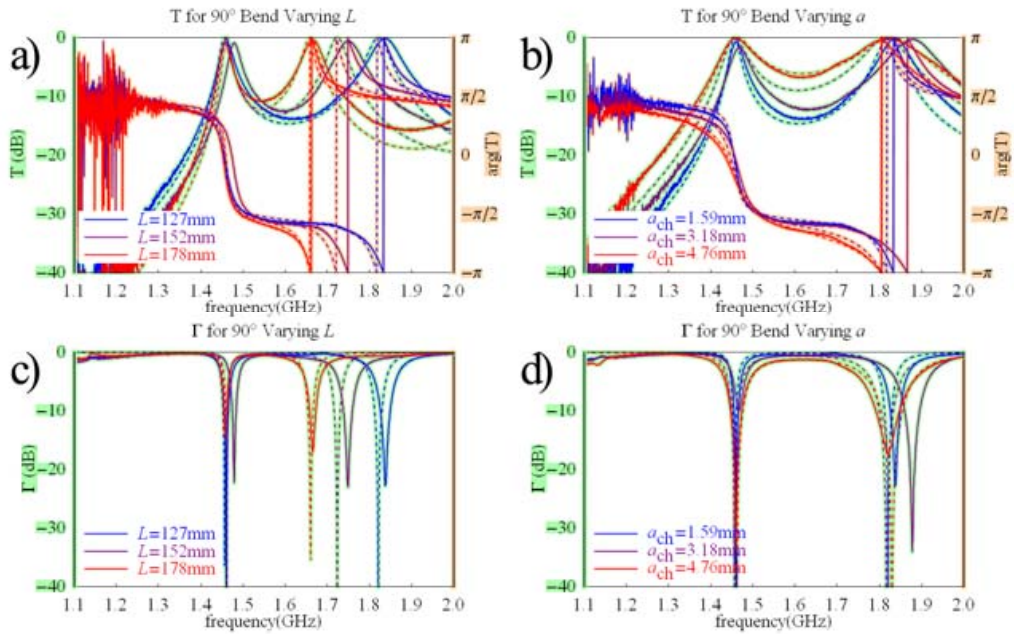

Figure 3: Similar to Fig. 2, but for the 90° bend configuration.

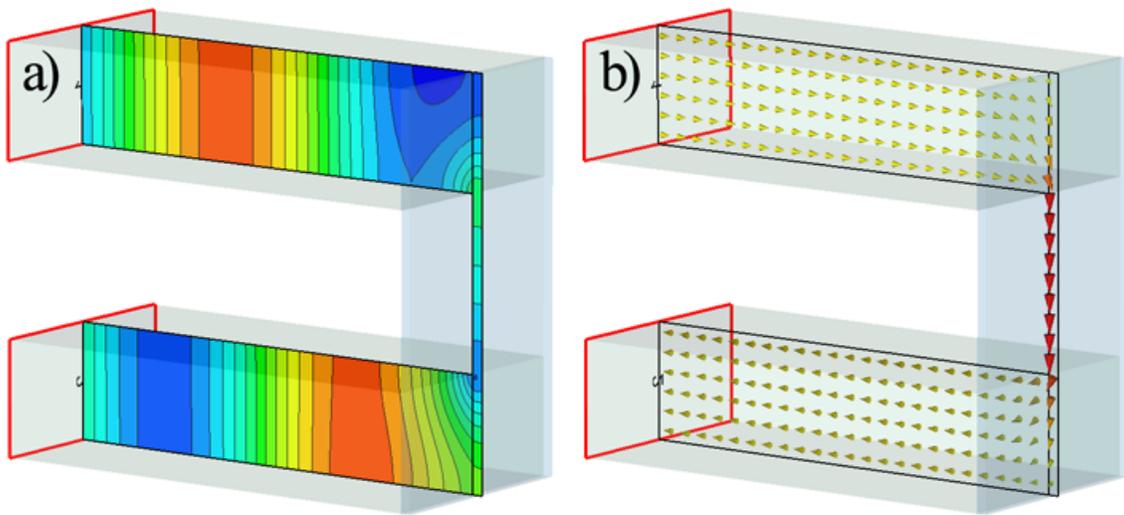

Figure 4: Simulations showing the fields internal to the 180° bend. Snapshot in time of the normal magnetic field (panel (a)) and real part of Poynting's vector (panel (b)). The channel width has been intentionally left relatively large for clarity, leading to a non-zero phase difference across the ENZ region. However, the fast-wave properties of the channel are very evident also in this configuration. Narrower

channels (as those experimentally realized and corresponding to Fig. 2-3) would provide an even smaller phase delay and faster wave propagation across the channel at the supercoupling frequency.

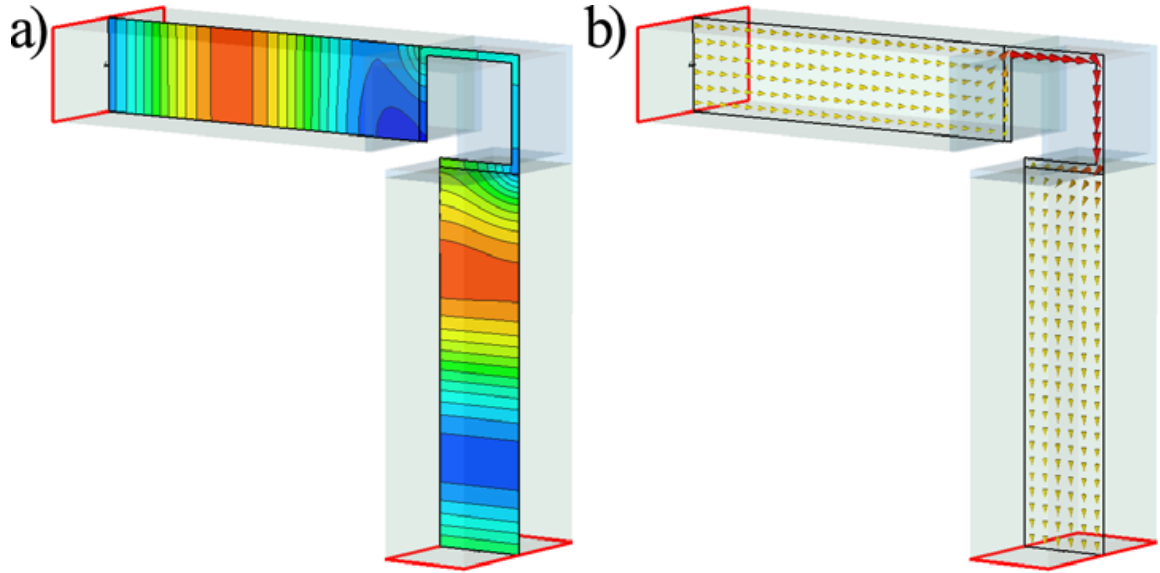

Figure 5: Similar to Fig. 4, but for the 90º bend.

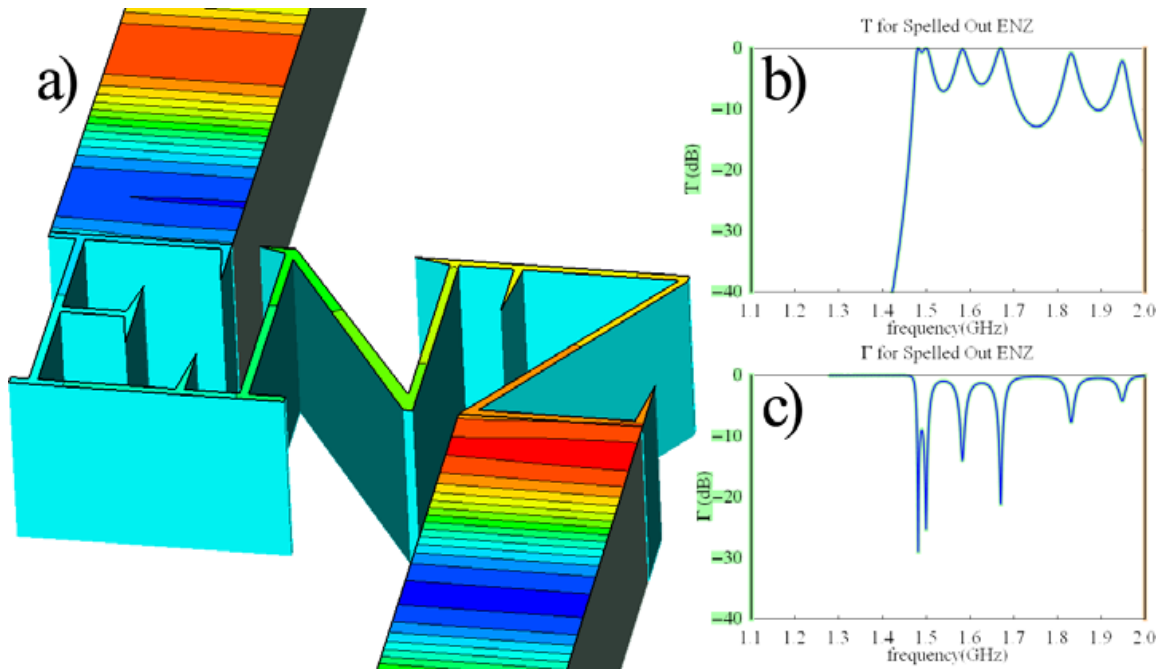

Figure 6: Simulation of the magnetic field distribution (panel (a), snapshot in time) in an ultranarrow waveguide channel spelling out the letters "ENZ" and connecting two much thicker waveguide sections. The plot confirms unity transmission and ultrafast wave propagation along the channel at the ENZ frequency, despite many turns, stubs, and approximately 2.5 free-space wavelengths of length at the cut-off frequency of the channel. The energy is squeezed into a channel approximately $1/20^{th}$ the original waveguide thickness. Transmission (panel (b)) and reflection (panel (c)) coefficients for the same geometry. The frequency relative to panel (a) corresponds to the first transmission peak.